\documentclass[conference]{IEEEtran}
\IEEEoverridecommandlockouts
\usepackage{cite}
\usepackage{amsmath,amssymb,amsfonts}
\usepackage{algorithmic}
\usepackage{graphicx}
\usepackage{textcomp}
\usepackage{xcolor}
\usepackage{amsmath}
\usepackage{hyperref}
\hypersetup{
    colorlinks=true,
    linkcolor=blue,
    filecolor=magenta,      
    urlcolor=cyan,
    pdftitle={Overleaf Example},
    pdfpagemode=FullScreen,
    }

\def\BibTeX{{\rm B\kern-.05em{\sc i\kern-.025em b}\kern-.08em
    T\kern-.1667em\lower.7ex\hbox{E}\kern-.125emX}}
\begin{document}

\title{Uber Stable: Formulating the Rideshare System as a Stable Matching Problem   \\
}

\author{

\IEEEauthorblockN{Rhea Acharya\textsuperscript{1}}
\IEEEauthorblockA{rlacharya@college.harvard.edu}
\and
\IEEEauthorblockN{Jessica Chen\textsuperscript{1}}
\IEEEauthorblockA{jessica\_chen@college.harvard.edu}
\and
\IEEEauthorblockN{Helen Xiao\textsuperscript{1}}
\IEEEauthorblockA{helenxiao@college.harvard.edu}
}

\maketitle

\begin{abstract}
Peer-to-peer ride-sharing platforms like Uber, Lyft, and DiDi have revolutionized the transportation industry and labor market. At its essence, these systems tackle the bipartite matching problem between two populations: riders and drivers.  This research paper comprises two main components: an initial literature review of existing ride-sharing platforms and efforts to enhance driver satisfaction, and the development of a novel algorithm implemented through simulation testing to allow us to make our own observations. The core algorithm utilized is the Gale-Shapley deferred acceptance algorithm, applied to a static matching problem over multiple time periods. In this simulation, we construct a preference-aware task assignment model, considering both overall revenue maximization and individual preference satisfaction. Specifically, the algorithm design incorporates factors such as passenger willingness-to-pay, driver preferences, and location attractiveness,  with an overarching goal of achieving equitable income distribution for drivers while maintaining overall system efficiency.

Through simulation, the paper compares the performance of the proposed algorithm with random matching and closest neighbor algorithms, looking at metrics such as total revenue, revenue per ride, and standard deviation to identify trends and impacts of shifting priorities. Additionally, the DA algorithm is compared to the Boston algorithm, and the paper explores the effect of prioritizing proximity to passengers versus distance from city center. Ultimately, the research underscores the importance of continued exploration in areas such as dynamic pricing models and additional modeling for unconventional driving times to further enhance the findings on the effectiveness and fairness of ride-sharing platforms.

Our simulation code can be found \href{https://github.com/rheaacharya11/uberstable}{here}.

\end{abstract}


\section{Introduction}
The subway breaks down on our way to work, it's 2:30 a.m. and we need to get home after a night of clubbing, we're in a new city and need to get from the airport to a museum across town — what's our first instinct? "Call an Uber." In this digital age, we rely on ride-sharing apps as an integral part of daily life. Passengers turn to driving apps, trusting that they will be able to get from their current destination to the next in a fast and safe way, without breaking their wallets. Ride-sharing apps like Uber, Lyft, and Didi, eager to gain more revenue, prioritize meeting these needs to provide a positive experience to passengers, hoping they will use the apps more and spend more money.

This means that apps have historically maximized rider satisfaction, ensuring that riders are likely to continue to use their services, as without them, the apps would not be able to survive. However, research has shown that in matching problems like this one, when the needs of one party is prioritized, those of the other can be sometimes hurt. Here, by prioritizing riders, the matching mechanisms might be creating pessimal outcomes for the drivers. If the negative effects are too large, this could have significant consequences on driver satisfaction and equity. 

Potential concerns related to this have already been seen by the public eye. In recent years, these ridesharing apps have come under fire for the driver experience that they provided. For example, a CNBC article found that only 4\% of Uber drivers are still driving a year later. Additionally, Uber and Lyft have bought been sued for withholding money from drivers, and drivers are frequently commenting on the unsustainability of the role in outlets like The New York Times. Why is this the case? Is it inherent to the nature of the role, or is it at all impacted by the Uber matching algorithm itself?

Motivated by these questions, we want to dive deeper into the issue of driver equity, exploring through the driver's perspective how the design of matching algorithms can affect the efficiency and equity. Our deliverable will have two main components: 1) a literature review of existing ridesharing platforms and efforts at increasing driver satisfaction; 2) a novel algorithm implementation with simulation testing. We will spend significant time defining an application of the Gale-Shapley deferred acceptance algorithm to this rideshare system matching problem, and we will compare the results of this to other algorithms, like random matching and closest neighbours.

\section{Previous Work}
 To begin, we want to better understand how the Uber system works and the general ride-sharing matching algorithm. Unfortunately, ridesharing services such as Uber and Lyft are not required to share data like taxi services, so exact algorithms can not be found through literature, yet we will review work discussing the current research of these algorithms and what they aim to achieve (in most detail). In order to ground our knowledge of current work in the field to inform the design of our simulation, we conducted a preliminary literature review. Additionally, we will focus on the problems that these algorithms pose to the driver-side of the system: in particular, we will do brief literature review on the trade-offs between maximizing objectives in ride-share matching algorithms and driver revenue distribution. Next, we will look at how Uber and other ride-sharing companies generate matching algorithms to best generate aggregate revenue for the company, and how this may affect consumers of the app. Ultimately, we will use these insights to inform the design of our simulation and will aim to work towards a solution to various problems addressed, through our preference ordering and nuanced matching algorithm.

\subsection{Two-Sided Fairness}

Most ride-sharing platforms prioritize maximizing passenger satisfaction, as this is directly correlated with maximizing revenue. There is a trade-off between income distribution and maximizing the overall objective, which in this case is to maximize revenue of app and minimize the total distance covered by drivers.  While existing platforms primarily prioritize customer satisfaction,  Sühr et al [1] emphasize the importance of considering fairness for drivers to ensure their well-being and long-term stability. The authors propose a framework for fair income distribution by amortizing equality over time, allowing drivers to receive benefits proportional to their active time on the platform. Specifically, the paper suggests algorithms that minimize income inequality while fairly distributing increases in customer waiting times. As we construct our own algorithm and analyze our results, we try to strike a balance between these two conflicting ideals, as we applied its concepts to a static matching problem rather than dynamic. [1]
\subsection{Preference-Aware Task Assignment}

Previous work in this field includes OSM-KIID [2], a model whose long-form name is Online Stable Matching under Known Identical Independent Distributions. In many matching problems, especially those constructed by commercial platforms like Uber, the aim is simply to maximize the profit of the system as a whole. This means that frequently, the individual preferences of parties on both sides are neglected, which in the short term might seem inconsequential but in the long term could have more drastic impacts, as participants may not longer feel that it is worthwhile to themselves to engage in the matching process. For example, in the rideshare system, if drivers feel as though their time and effort are not being compensated appropriately or they always get the least desired rides (like those that start or end in remote areas), then they may quit their job which is not good for the system as a whole. The authors of OSM-KIID seek to create a new model that takes into account both the overall revenue minimization objective as well as the preference satisfaction one, by maximizing revenue and minimizing blocking pairs. We incorporate this focus on preferences within our deferred acceptance algorithm, by taking into account factors like prior income distribution and desirability of starting and ending locations within the preference ordering generation. Additionally, we account for individual diversity of preferences by having unique cost coefficients among other nuances. [2]

\section{Algorithm Design}
\subsection{Overview}
At the most fundamental level, we want to model the rideshare system as a two-sided matching problem. Then, we will explore how we can design our matching algorithm to this bipartite problem with driver equity in mind. Specifically, we want to achieve equitable income distribution among drivers while 
maintaining comparable total welfare, as measured 
by total revenue and aggregate wait time. Although, in reality, the algorithms used by Uber, Lyft, and other ride-sharing apps are dynamic, given the shortened timeframe of this project, we will limit our algorithm to a static algorithm run over multiple time periods.

\subsection{Approach}
\begin{enumerate}
    \item Reformulate this problem into a deferred-acceptance two-sided matching problem.
    \item Driver-proposing: Create cost functions for drivers and (fixed) payment functions. Drivers will propose a ride if utility is positive for a given request.
    \item Both drivers and riders have preference orders, and of note is the fact that we will incorporate driver income into preference ordering to look into how driver equity is prioritized
\end{enumerate}

\subsection{Some Relevant Assumptions}
\begin{itemize}
\item We do not have equilibrium pricing. Instead, price is determined by riders’ fixed willingness to pay given location with their WTP is fixed and known in a given round.
\item We constrain goal to driver-side only, so we assume riders simply want to get picked up.
\item We have our ride-sharing platform as a benevolently matching central allocation mechanism, with no revenue generation abilities.
\item We model this as a static problem, where the algorithm only aims to solve problem for all drivers and riders currently on the map. Drivers are constant through all the rounds, in order to allow us to most helpfully keep track of driver income, but we can generate new passengers at each round. 
\end{itemize}

\subsection{Algorithm Flow}

The Gale-Shapley algorithm can be applied to this situation as follows:
\begin{enumerate}
    \item Each rider has a willingness-to-pay(WTP) and both a pickup and drop-off location.
    \item Next, drivers see if utility, deemed to be price minus cost, is positive, then proposes rides to riders based on their preference orderings
    \item Rider decides to accept/reject depending on wait time, and will do so in an order based on their own preference orderings. The preference orderings for both drivers and riders will take into consideration certain heuristics/tradeoffs as detailed below.
\end{enumerate}
\subsection{Algorithm Discussion}
The Gale-Shapley algorithm, also known as deferred acceptance, results in a stable matching. This means that there is no blocking pair consisting of a passenger and driver who would prefer to be matched with each other than their current pairing, based on their preference ordering. Additionally, the Gale-Shapley algorithm is strategy-proof and optimal for the proposing side, which in this case consists of the drivers. This means that it is in the driver's best interest to report their honest attributes and preferences. Typically, the Gale-Shapley algorithm is not strategy-proof for the non-proposing side, and so there can be useful deviations for those entities. However, some concerns here are less relevant in this specific case, given the name of ridesharing requests. Since riders are currently in one specific location and want to get to another, then it would not make sense to misreport either of these values as then 

The area of greater concern could be in the underreporting of willingness-to-pay (WTP) by passengers, as if they are able to be still matched with this lower WTP then they would prefer to pay this lower amount if they are still completing their ride, and thus having the same utility. We could potentially fix this by having the payment amount be a critical value rather than a fixed WTP, such that below this point, the ride would not be completed and the passenger would not derive any utility.

We can make our algorithm more equitable if drivers care less about the drop-off location's distance from the center. If drivers are willing to prioritize passengers that are farther away, then this would help overall driver income equity but it would hurt the company's overall revenue.
\subsection{Utilities Functions}
Each passenger has a fixed willingness-to-pay (WTP), calculated as a linear function of the Manhattan ride distance and the Euclidean distance from city center, wrapped in a Gaussian distribution. In other words, 
\begin{align*}
    WTP = \alpha ((\mathtt{rideLength} + \mathtt{cityCenterDist}))/3
\end{align*}
where $\alpha$ is a coefficient randomly sampled from a Gaussian distribution with mean = 1 and standard deviation = 0.1, $\mathtt{rideLength}$ is the Manhattan distance from the passenger's requested pickup and dropoff locations, and $ \mathtt{cityCenterDist}$ is the mean of the Euclidean distances of the pickup and dropoff locations from the city center. The stochasticity of $\alpha$ denotes that different riders may have different tradeoffs and demand factors that are outside of our current scope, naively capturing different people's frugality. We calculated the Passenger willingness ot pay like this to demonstrate that passengers are willing to pay more for longer rides, as well as rides that make the driver go more out of the way (rides that begin and end farther from the city center).

Similarly, each driver's cost is a function of their total trip length and distance from the city center, multiplied by a personal cost coefficient.
\begin{align*}
    \mathtt{cost} = \gamma (\mathtt{tripLength} + \mathtt{cityCenterDist})
\end{align*}
where $\gamma \in [0,1]$ is a cost coefficient randomly sampled from a Gaussian distribution, $\mathtt{rideLength}$ is the sum of the Manhattan distances from the driver's current location to the passenger pickup location and the passenger's pickup to dropoff locations, and $ \mathtt{cityCenterDist}$ is the Euclidean distance of the dropoff locations from the city center. The cost coefficient of $\gamma$ denotes that different drivers may have cost factors that are outside of our current scope. 

While our riders simply post their WTP, our drivers must take into consideration the difference between a rider's WTP and the cost of the trip. In other words, we have simplified market equilibrium dynamics to be constrained to just driver-side profit and utility. Our drivers have higher costs for longer drives, as well as drives that result in them being father away from the city center that would limit their ability to pick up new rides.

\subsection{Preference Orders}

Both drivers and passengers take into consideration multiple weighted factors when determining their preference orders. In our project, we experiment with these hyperparameters for how much weight should be placed on these different factors to achieve the best results. Here, lower preference scores are prioritized.

Drivers take into account two factors: profit and distance from the city center. Normalizing either factor, driver $d$ calculates the following preference score for a passenger $p$:
\begin{align*}
    pref_d (p) = & -\mathtt{p.WTP}+ w_p \mathtt{d.cost(p)} \\ 
    \mathtt{d.cost(p)} = & \alpha (\mathtt{d.pickupDist(p)}) \\
    &+ 0.1*(\mathtt{d.cityCenterDist(p)})^2 \\ 
\end{align*}

where $w_p$ is a randomly generated gaussian random variable with mean = 1 and SD = 0.2 to demonstrate that different drivers value cost of the drive different from each other. $\mathtt{d.pickupDist(p)}$ is the manhattan distance between the driver and passenger's pickup location and $\mathtt{d.cityCenterDist(p)}$ is the euclidean distance between the passenger's dropoff location and the city center, normalized. Drivers prioritize rides that earn them more money, which is why we add the WTP of the passenger (the whole function is flipped by the negative since sorting in python sorts smallest to largest), while they also want rides with the smallest cost to them possible.

Passengers also take into account two factors when ordering drivers: wait time and the driver's current aggregate income. This is where our system attempts to implement income equity among the drivers. By adding this factor into the preference orders, the system allows drivers with lower aggregate income in the current period to get higher priority in matching, thus hypothetically promoting greater ride equity. We will test whether this is a plausible solution or not in the simulation. Normalizing either factor, passenger $p$ calculates the following preference score for a passenger $d$:
\begin{align*}
    pref_p (d) = &w_t (\mathtt{waitLength(p,d)}) \\ 
     &+ w_i (\mathtt{aggIncome(d)})
\end{align*}
where $w_t$ is the weight applied to the pickup wait time (which, due to our unit time per distance, is just the Manhattan distance between the driver's current location and the passenger's requested pickup location) and $w_i$ is the weight applied to the driver's aggregate income thus far. When the system does not take into consider income distribution, we set $w_t = 1$ and $w_i = 0$. We hypothesize that the greater the $w_i$, the more equitable our overall income distribution, at the tradeoff of our aggregate system revenue.

\section{Simulation Design}
\subsection{Code}
The Github repository containing our original simulation code can be found \href{https://github.com/rheaacharya11/uberstable}{here}.
\subsection{City Framework} We will simulate our city using a 100 x 100 Euclidean Grid, with the idea being that it takes 1 time unit to travel a single Euclidean distance. An important characteristic of many cities is the city center, which will assume to be the direct center of the grid. This is the heart of the city where most people and places of interest exist. 

Thus, we generated locations to be Euclidean points where the x,y coordinates span from -50 to 50. To account for the population densities, they are simulated randomly from a normal distribution (with mean = 0 and standard deviation = 20) to simulate that there are more people traveling to and from the city center.

This is not only relevant for generating what the initial function of each passenger and driver is, but it is only relevant because then drivers will prefer rides that have drop-off locations closer to the city center rather than those that end up in more remote locations. This is because then they will be more likely to be closer to their own location of interest or to a new ride.

\subsection{Driver Generation}
Each driver will be instantiated at the start of the first round. They will have the following attributes, each initialized to 0, None, or the empty list: $\mathtt{total income}$, $\mathtt{current location}$, $\mathtt{total rides}$, $\mathtt{preference ordering}$, $\mathtt{matched passenger name}$, and $\mathtt{locations}$, which is a list of locations that they have visited. With each driver, we will also generate a unique cost coefficient and a starting location, determined through the function of Euclidean distances from the center touched upon above. Our set of drivers will stay constant throughout the rounds,  so that we can best analyze the effects on cumulative driver income equity. In each subsequent round, we will update the attributes intuitively based on the matching and completed ride. The cost coefficient will stay the same for a particular driver between rounds and the new starting location will just be the ending location of the previous ride if one was completed.

\subsection{Passenger Generation}
Unlike drivers, we will generate a new set of passengers in each round. Each will have the following attributes, initialized to 0, None, or the empty list as best fits: $\mathtt{preference ordering}$, $\mathtt{WTP}$, $\mathtt{current location}$, and $\mathtt{dropoff location}$. Once the matching and rides has been completed, we assume the passengers are no longer involved in the next round and generate an entirely new batch, so no updating is necessary directly to the passengers.

\subsection{Design}

For the sake of having enough participants but keeping our computational time reasonable, we chose to have 15 drivers and 15 passengers in any given round, and we iterated over 50 total rounds to see the long-term distribution of income. In addition to specifying the number of rounds and the number of agents per round, the simulation can also vary the Gaussian distribution of passenger instantiation away from the city center, standard deviations of driver cost functions and passenger WTPs, as well as the weight each agent may assign its different priorities.

\subsection{Random Matching} 
As a baseline, we explore what occurs when passengers and drivers are matched randomly, which we implement by randomly shuffling both lists and then creating matches in order.

\subsection{Closest Neighbours}
Next, we will apply an algorithm that first matches the pairing that are closest together. More concretely, we implement this by constructing a Euclidean distance matrix between passengers and drivers and then solving the matching problem using a Hungarian algorithm. In this case, this is a central allocation mechanism, in that it is neither rider-proposing nor driver-proposing and instead Uber is acting as the third party to create the matches.

\subsection{Gale-Shapley}
The heart of our exploration focuses on the results that occur when we use the Gale-Shapley deferred acceptance algorithm. 

\subsection{Altering Weights}
Within this Gale-Shapley algorithm, we encapsulate various factors, such as Manhattan distance, cost, distance from city center, and driver income distribution, into one equation under preference orderings for passengers' ranking of drivers. In this process, we assign different weights to these factors, corresponding to how much we care about each factor. Part of our results will  be in modifying these weights and seeing how this leads to different overall distributions and what tradeoffs surface.

\section{Results}
For the most thorough understanding of the performance of the deferred acceptance algorithm, we will compare the results against that of Random Matching and Closest Neighbors algorithms. First, we will output a chart of results of the different algorithms against each other. First, let us look at the total cumulative incomes of the four different algorithms

\begin{center}
\begin{tabular}{||c c c c||} 
 \hline
 DA Driver prop & Random & Closest & Boston Algo \\ [0.5ex] 
 \hline\hline
 14644.953&	10084.181	& 12839.953 &	8196.113 \\
  \hline
15934.399	&9555.306 &	12956.324 &	11372.515\\
 \hline
17567.636	&9247.306 &	13789.982 &	7963.459\\
 \hline
18106.353	&10164.92 &	13058.475 &	12930.511\\
 \hline
17168.541 &	9536.810 &	14708.658 &	12343.745\\
 \hline
17131.983	& 9469.655 &	12671.108 &	11733.939\\
 \hline
13544.27 &	9469.239	& 11482.561 &	9644.839\\
 \hline
16419.518 &	8971.500 &	12421.540	& 12778.096\\
 \hline
15864.458	& 10368.144 &	13716.894 &	12983.909\\ [1ex] 
 \hline
\end{tabular}
\end{center}

As we can see from the chart, the driver proposing DA algorithm always produced the greatest aggregate income, followed by Closest (closets drivers and passengers are matched without any side proposing), then the Boston algorithm where drivers simply propose without passengers able to switch, followed by random. This makes sense since drives are rejected by passengers when wait time is too long, which results a lot more often in random and Boston than Closest or DA.

\subsection{Graph comparing Different Algorithms}
These graphs plot the title quantity in terms of number of agents (total drivers plus passengers) in a given time period.

\begin{figure}[h]
\begin{center}

    \includegraphics[scale = 0.35]{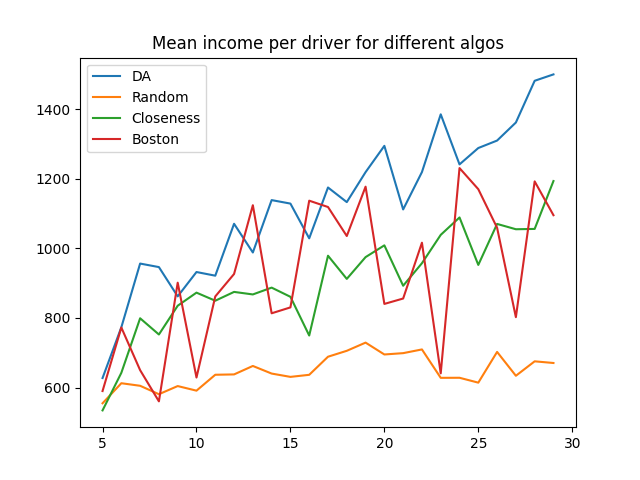}
    \vspace{-20pt}
    \caption{\textit{Mean income per driver vs. number of agents.} In the randomly matched algorithm, we see that mean income stays around the same at \$600 per driver, which is as expected, since the ratio of drivers and passengers stay 1-to-1. In contrast, we see that the Boston matching algorithm demonstrates greater variability with increased number of agents, since given the instability, there is increased capacity for the algorithm to output suboptimal matchings between drivers and passengers, where passengers may not give input on better wait times, which may imply better allocation. Then, we see that the closeness and DA matching algorithms are better equipped to capture greater revenue as agents increase. These results demonstrate that proximity is a driver of revenue, and the DA algorithm builds on this revenue driver by further considering of profit and city-center-clustering.}
    \vspace{-20pt}
\end{center}
\end{figure}

\begin{figure}[hbt!]
\begin{center}
    \includegraphics[scale = 0.35]{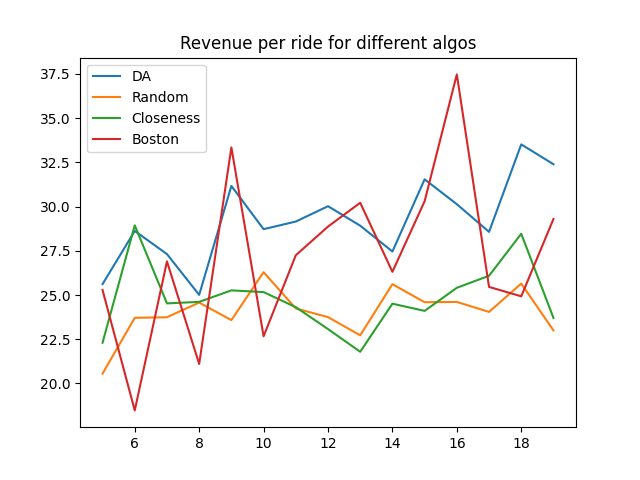}
    \vspace{-20pt}
    \caption{\textit{Revenue per ride vs. number of agents.} Here, we see that, despite significantly better revenue performance for closeness matching and DA, we see that the the closeness matching and random matching algorithms maintain relatively similar per-ride revenue.}
    \vspace{-20pt}
\end{center}
\end{figure}

\begin{figure}[h]
\begin{center}

    \includegraphics[scale = 0.35]{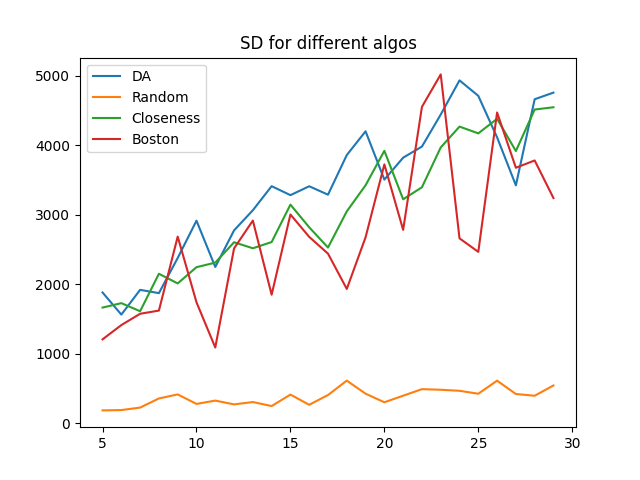}
    \vspace{-20pt}
    \caption{\textit{Standard deviation of income per driver vs. number of agents.} As expected, standard deviation of driver income of the non-random algorithms increase with the increased number of agents, while the random algorithms maintains around the same variability.}
    \vspace{-20pt}
\end{center}
\end{figure}

\begin{figure}[h]
\begin{center}
    \includegraphics[scale = 0.3]{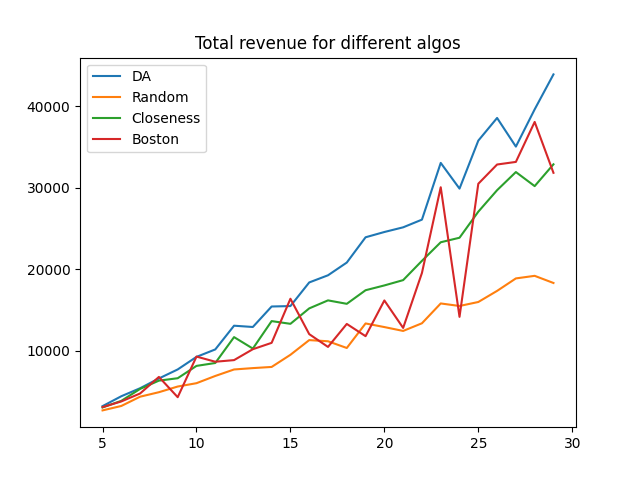}
    \vspace{-20pt}
    \caption{\textit{Total revenue of system vs. number of agents.}  This graph further supports our findings in previous figures, which demonstrate that the DA algorithm best capitalizes on more agents' preferences to maximize revenue for drivers. We also continue to see that the Boston matching algorithm demonstrates higher variability due to instability.}
    \vspace{-20pt}
\end{center}
\end{figure}

\subsection{Graph comparing DA with Fairness}
\begin{figure}[hbt!]
\begin{center}
   \includegraphics[scale = 0.35]{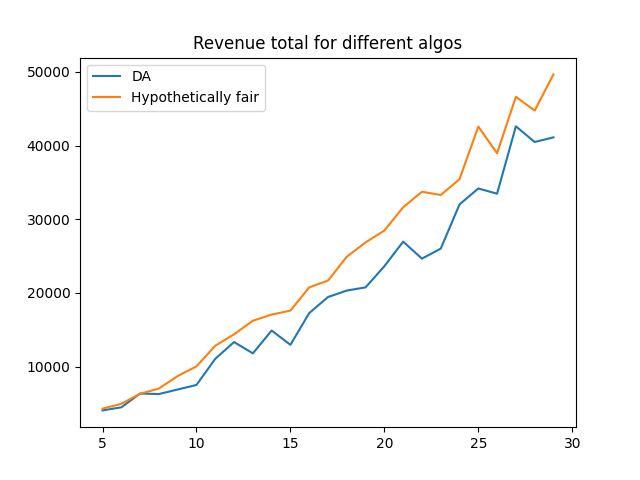}
   \vspace{-20pt}
   \caption{\textit{Total revenue based on fairness vs. number of agents.} Interestingly, we find that our hypothetically fair algorithm (DA with driver income taken into consideration for preference ordering) achieves higher aggregate revenue compared to the vanilla DA algorithm. This result would suggest that equitable income distribution not only has positive implications on the labor market and sustainable business practices, but also increases welfare and revenue generation. Still, we may have to do further work to confirm the drivers for this revenue-driving equity.}
   \vspace{-20pt}
\end{center}
\end{figure}

\begin{figure}[!ht]
\begin{center}
   \includegraphics[scale = 0.35]{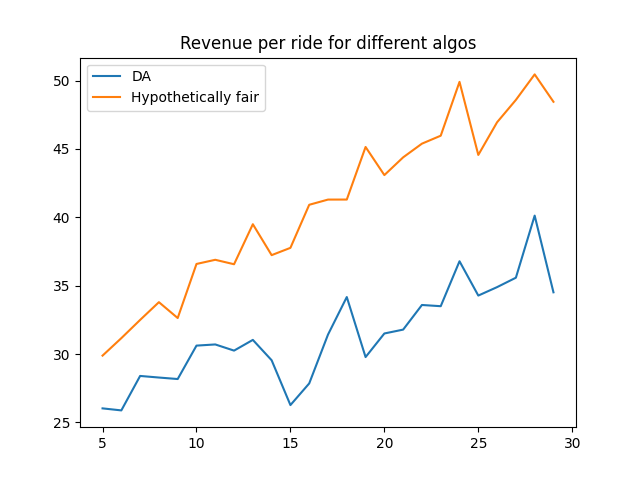}
   \vspace{-20pt}
   \caption{\textit{Revenue per ride based on fairness vs. number of agents.} Again, we see our hypothetically fair algorithm performing better in increasing revenue per ride.}
   \vspace{-20pt}
\end{center}
\end{figure}

 \begin{figure}[h]
 \begin{center}
    \includegraphics[scale = 0.35]{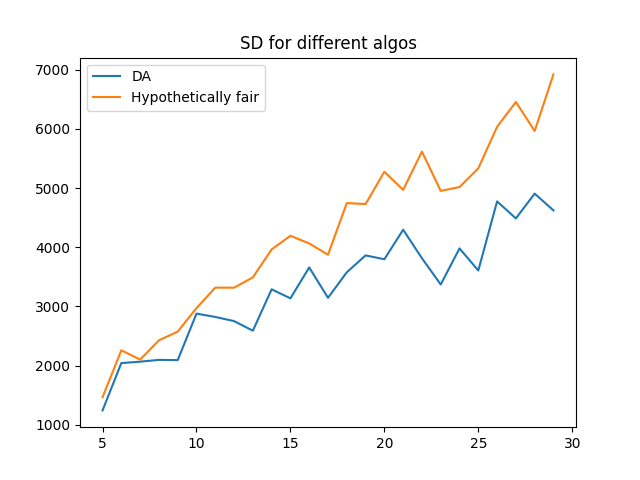}

     \caption{\textit{SD based on fairness vs. number of agents.} The hypothetically fair algorithm sees slightly greater standard deviation in income per driver. }

\end{center}
\end{figure}

\subsection{Graphs with new fairness weights}

\begin{figure}[t]
\begin{center}
    \includegraphics[scale = 0.3]{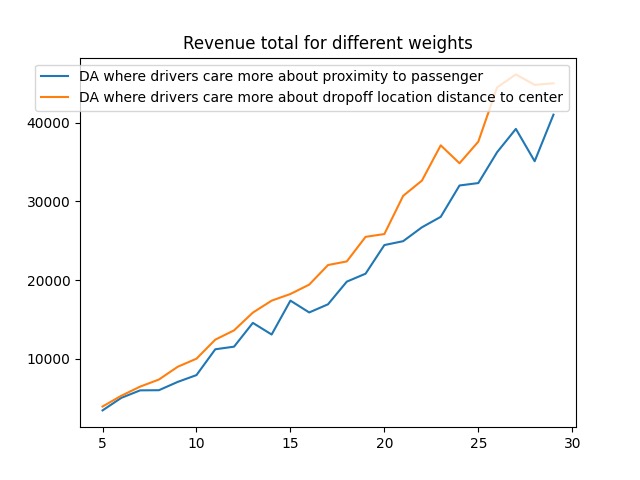}
    \caption{\textit{Revenue total vs. number of agents.} Drivers better optimize revenue by prioritizing an optimal dropoff location (one closer to the city center, where they are likely to book the next passenger with less travel cost) over naive proximity to passenger. This result suggests that the clustering of drivers around passenger hot spots is significant in generating greater revenue. Therefore, in order to incentivize drivers to drop off passengers in more remote locations, the system may have to take further measures at distributing income.}
\end{center}
\end{figure}
\begin{figure}[t]
\begin{center}
    \includegraphics[scale = 0.3]{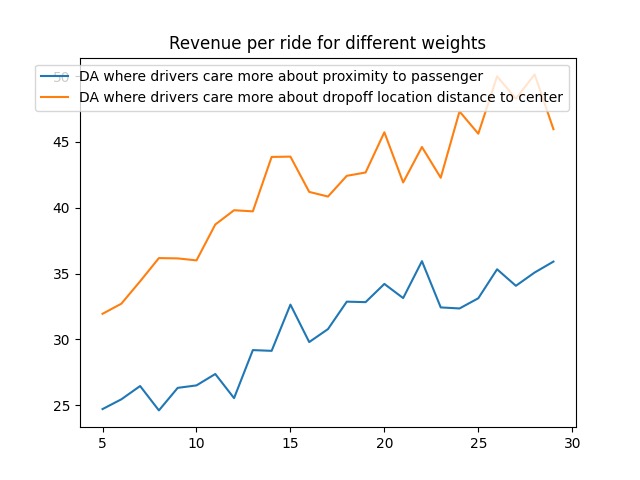}
    \caption{\textit{Revenue per ride vs. number of agents} This graph supports the one above, showing a strong difference between the revenue per ride for different weights. In particular, total revenu is increased when drivers care more about the dropoff location's distance from the center, but this might be at the tradeoff of less equity across drivers.}
\end{center}
\end{figure}
\begin{figure}[h]
\begin{center}
    \includegraphics[scale = 0.3]{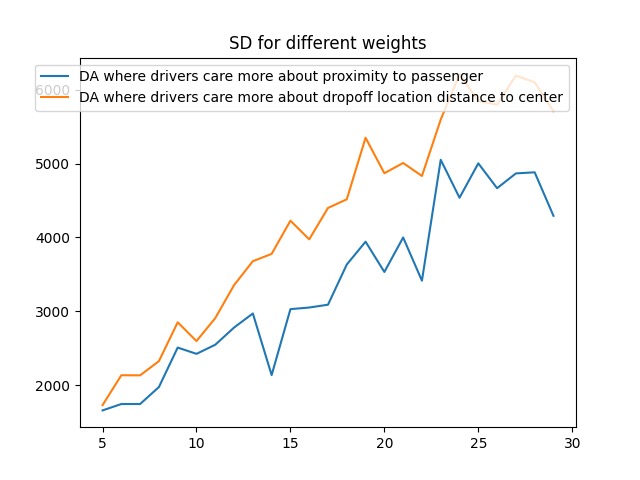}
    \caption{\textit{Standard deviation vs. number of agents} As we see, the standard deviation of the driver incomes where drivers put more weight onto dropoff location of passengers is higher than the sandard deviation of the driver incomes that put more weight in passenger proximity. This demonstrates that putting more focus on reaching driver with farther locations to travel can bring more income equity, lowing the variance of driver incomes and flattening driver income distribution.}
\end{center}
\end{figure}

\newpage 

\section{Reflections and Conclusion}

In conclusion, we found that the Gale-Shapley deferred acceptance (DA) algorithm was better equipped to optimize driver revenue than other matching algorithms, such as closeness matching or the Boston algorithm. We also found that stability is important for better income equity, as the Boston algorithm showed high variability for most metrics of results. 

Surprisingly, we found that our hypothetically fair algorithm (which is the DA algorithm modified to prioritize drivers with lower aggregate income) was able to generate higher revenue than the vanilla DA algorithm. This may imply that pursuing income equitability is not only a labor-sustainable practice, but also something that may achieve revenue upsides.

The ridesharing system is complex and hard to comprehensively evaluate in any one research endeavor, much less one for an undergraduate class. As such, there are several further points of study. Of interest to us is implementing a dynamic or competitive pricing model, where prices increase if demand is greater than the supply, such as through origin-based surge pricing. Additionally, we would like to model behavior at abnormal times of driving, such as in the later hours of the night or in rush hours like post-work or around holidays like Thanksgiving. This could give insights into how changes in demand and driver density could affect driver equity and treatment.

\section{Acknowledgements}



\end{document}